\documentclass{elsart}

\usepackage{natbib}
\usepackage{amssymb}
\usepackage{amsmath}
\usepackage{graphicx}
\usepackage{color}

\def\inbar{\,\vrule height1.5ex width.4pt depth0pt}
\def\IR{\relax{\rm I\kern-.18em R}}
\def\IC{\relax\hbox{$\inbar\kern-.3em{\rm C}$}}

\newcommand{\ket}[1]{|{#1}\rangle}
\newcommand{\bra}[1]{\langle{#1}|}                     

\def\Rb87{$^{87}\text{Rb}$}
\def\0{\ket{0}}
\def\1{\ket{1}}

\def \expect#1{{\left \langle #1 \right\rangle}}

\newcommand \bfr{{\bf r}}

\def\be{\begin{eqnarray}}
\def\ee{\end{eqnarray}}
\newcommand \bse{\begin{subequations}}
\newcommand \ese{\end{subequations}}
\newcommand \bml{\begin{subequations}\begin{eqnarray}}
\newcommand \eml{\end{eqnarray}\end{subequations}}
\newcommand \bfepsilon{{\boldsymbol \epsilon}}

\def\atomic#1{\tilde#1}
\begin{document}

\begin{frontmatter}



\title{Entanglement of Two Atoms usingÊ\\ Rydberg Blockade}


\author{Thad G. Walker and Mark Saffman}

\address{Department of Physics, University of Wisconsin-Madison, Madison, WI 53706}

\begin{abstract}
Over the past few years we have built  an apparatus to demonstrate the entanglement of neutral Rb atoms at optically resolvable distances using the strong interactions between Rydberg atoms.  Here we review the basic physics involved in this process:  loading of single atoms into individual traps, state initialization, state readout, single atom rotations, blockade-mediated manipulation of Rydberg atoms, and demonstration of entanglement. 
\end{abstract}

\end{frontmatter}


\section{Introduction}

One of the most rapidly progressing areas of scientific research over the past 25 years is the field of quantum information.  The possibility, once remote, of using entanglement of quantum systems as a resource, whether for quantum simulation, quantum computation, or enhanced quantum measurements, is rapidly being realized in research laboratories around the world.  While different workers in the field may share a variety of motivations for pursuing this work, it is safe to say that they share a fascination with studying and exploiting the strangest consequences of the quantum world.

A partial list of quantum systems that are being pursued includes trapped ions \citep{Blatt2008}, superconductors \citep{Clarke2008}, linear optics \citep{Kok2007},
semiconductor quantum dots \citep{Morton2011,Yamamoto2011}, nitrogen vacancy centers in diamond \citep{Togan2011}, and neutral atoms \citep{Zoller2011}.  For a physical system to be viable in the long run, it must exhibit most or all of the following elements  \citep{DiVincenzo1998}: well-defined qubits that allow for  initialization into well-characterized, long-lived quantum states with the capability of high-fidelity state-dependent readout;  a means to deterministically and controllably entangle individual qubits without decoherence; and the ability to transfer entanglement remotely.  While specific applications may relax certain of these properties, in general the above list describes the needs for generic quantum information processing.

We see neutral atoms as an attractive general system for manipulation  of quantum information.  They are similar to ions, the best developed system to date, in that they have long-lived hyperfine states that are robust against decoherence, and they can be precisely manipulated by optical and other electromagnetic fields. They may have some practical advantages vis-a-vis scaling to large arrays of qubits. {Independent of its promise for practical quantum manipulation, a fascinating and novel method for generating entanglement using Rydberg atoms directly motived the work described herein and is a topic of fundamental interest for atomic, molecular, and optical physics \citep{Saffman2010}.}

In 1999, \cite{Jaksch2000} proposed the concept of dipole or Rydberg blockade for entangling atom pairs at distances of greater than 1 micron, large enough for individual interrogation of the atoms by visible optical fields.  The basic idea, described in more detail in Section~\ref{sec:blockade}, is that the excitation of one atom into a Rydberg state will shift the energies of the corresponding Rydberg states of nearby atoms by more than the excitation linewidth.  This means that the quantum evolution of one atom can be controlled by the quantum state of another atom that may be many microns away.  Jaksch et al argued that this process could be done coherently and with high fidelity, and thereby could form the basis of a practical means of entanglement of neutral atoms.  Our critical evaluation \citep{Saffman2005a} agreed with this conclusion and outlined many of the experimental  and atomic physics issues facing its implementation in the laboratory.

This paper is a description of our experimental program to implement and investigate the Rydberg blockade entanglement mechanism in a two-atom system.  The first steps, described in Section~\ref{sec:trapping}, are to trap and cool pairs of atoms in separate far-off-resonant traps (FORTs),  to non-destructively measure their presence, and to read out their quantum states at the end of each experimental realization. 

Preparation of an arbitrary initial state for a computation requires optical pumping into one of the qubit basis states, followed by creation of single-atom superposition states using addressable stimulated Raman scattering (Section \ref{sec:stateprep}).  In addition to state preparation, these Raman rotations constitute a central feature of virtually any type of quantum gate that might be realized with the blockade mechanism.

While the readout and Raman manipulation capabilities borrow strongly from pioneering ion trap work \citep{Wineland1998},  the Rydberg blockade mechanism requires the capability of coherent resonant Rabi excitation and de-excitation of Rydberg states approaching principal quantum numbers $n=100$.  Section~\ref{sec:RydRab} describes our approach to this using two-photon excitation.

Once one atom has been excited to a Rydberg level, its ability to control its neighbor depends on the properties of the interactions between Rydberg atom pairs.  The figure of merit for blockade effectiveness, described in Section~\ref{sec:blockadeexpt}, is a weighted average of the inverse square of the interaction strength.  The need for {a strong blockade} drove us toward $n=100$, where the interaction strength is sufficient to provide effective blockade at 10 $\mu$m atom separations.

Section~\ref{sec:CNOT} describes the first implementation of a neutral atom CNOT gate.  The CNOT truth table was measured and it was verified that the basic physical mechanism of Rydberg blockade allowed the target to be transferred from state to state based on the internal state of the control atom.  The verification that the process preserves quantum coherence is described in Section~\ref{sec:parity}.

{In Section~\ref{sec:concl} we  discuss of some important factors needed for future improvements in the entanglement generated by Rydberg blockade.  The two major points of emphasis are the need for deterministic loading, and a number of important advantages of dark state FORTs.}

We would be remiss not to note that many fascinating Rydberg-related experiments have been done in extended cold atom samples. These have been { reviewed} recently by \cite{Choi2007}, \cite{Comparat2010}, and \cite{Saffman2010}.

\section{Entanglement Using Rydberg Blockade}\label{sec:blockade}

\begin{figure}
\includegraphics[width=3.2 in]{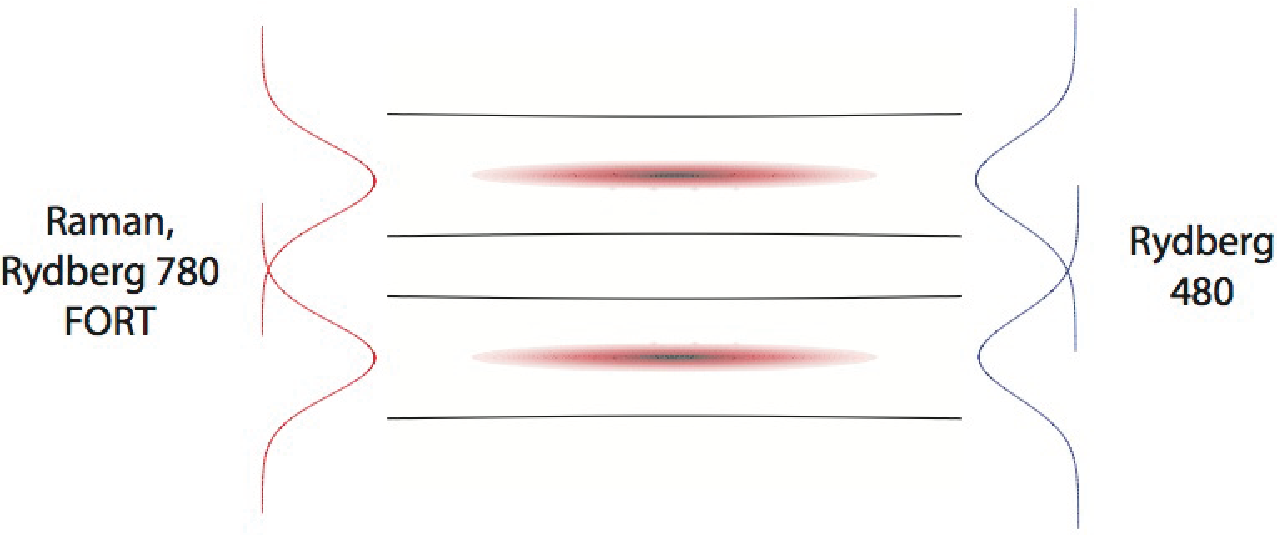}\hspace*{0.2 in}
\includegraphics[width=1.7 in]{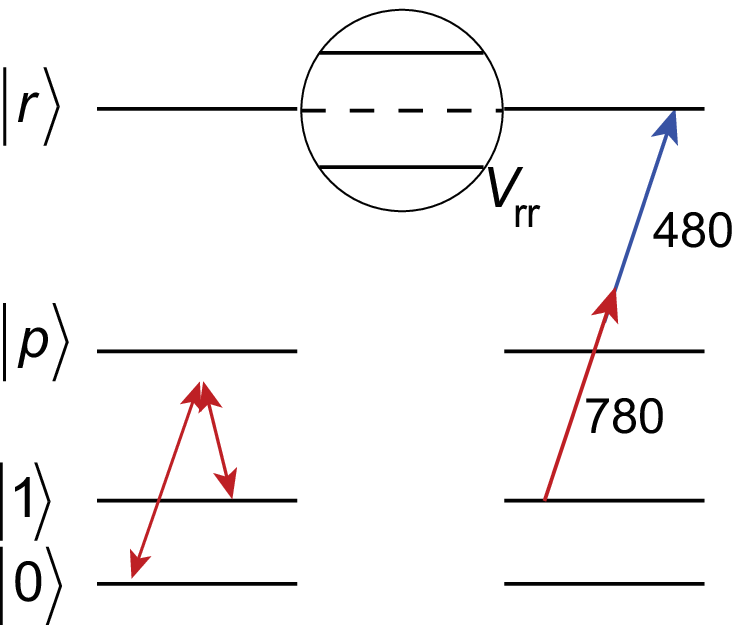}
\caption{(left) Geometry of Rydberg blockade. Two atoms are trapped in separate FORTs.  Each atom can be addressed with Raman and Rydberg resonant lasers to perform single-atom and two-atom gates.  (right) Energy levels and laser wavelengths.  The qubit states $\ket{0}$ and $\ket{1}$ are two hyperfine levels of the 5s ground state of Rb.  Raman and Rydberg Rabi excitations are driven by 2-photon resonant lasers through the virtual $\ket{5p}$ excited state. The Rydberg-Rydberg interaction $V_{rr}$, present when both atoms are in the Rydberg state $\ket{r}$, 
{enables entanglement by blocking excitation of both atoms to the Rydberg state.}}\label{blockadegeometry}
\end{figure}

The geometry for quantum information processing using the Rydberg blockade concept is illustrated in Fig.~\ref{blockadegeometry}.  Two atoms are trapped by a pair of focussed laser beams separated by a distance $R$ on the order of 10 $\mu$m.  The atoms have two hyperfine levels, labeled $\ket{0}$ and $\ket{1}$, that represent the qubit basis states. Each atom can be manipulated by moveable and switchable laser beams.  The first set of beams perform stimulated Raman transitions between the qubit basis states, and allow each qubit to be placed in an arbitrary superposition of $\ket{0}$ and $\ket{1}$.  A second set of lasers perform coherent 2-photon excitation of state $\ket{1}$ to a Rydberg level $\ket{r}$ and back.   Using these capabilities we {can, in principle, place} each atom in arbitrary superpositions of the states $\ket{0}$, $\ket{1}$, and $\ket{r}$.

The key physics behind the blockade process concerns the interactions between the two atoms in the various states.  When both atoms are in the qubit states, the dominant interatomic interaction is the magnetic dipole dipole interaction which is of order 
\be
V_{qq}\sim{\mu_B^2\over R^3}\sim 10^{-5}  \mbox{Hz @R=10 $\mu$m}
\ee
where $\mu_B$ is the Bohr magneton and $R$ is the distance between the atoms. This interaction is completely negligible; at these distances, atoms in neighboring FORTs do not interact with each other on the time scale of these experiments.

The interaction between a Rydberg atom of principal quantum number $n$ and a ground state atom of effective principle quantum number $n_g$ is dominated by the second-order dipole-dipole interaction, and is of approximate magnitude
\be
V_{qr}={(n^2e a_0)^2(5 e a_0)^2\over E_{sp} R^6}\sim 10^{-5}  \mbox{Hz @R=10 $\mu$m}
\ee
Here $E_{sp}\sim {\rm Ry}/n_g^3$ is a typical excitation energy for the ground state atom, { $e$ is the electron charge, $a_0$ is the Bohr radius,} and the two terms in the numerator are the {approximate  magnitudes of the dipole moments} for the Rydberg and ground-state atoms.  Again, this interaction is completely negligible.

In contrast, the interaction between two Rydberg atoms often approaches the resonant dipole-dipole limit  \citep{Walker2008}:
\be
V_{rr}={(n^2e a_0)^2\over R^3}\sim {n^4\over \atomic{R}^3}\sim 100   \mbox{MHz @R=10 $\mu$m},
\ee
much larger than the natural line width of typical Rydberg states and significantly larger than potential light-atom couplings of a few MHz for MHz quantum logic rates.  Thus if we put one of the trapped atoms into a Rydberg state using resonant light, the corresponding Rydberg state for the other atom will be highly off-resonant and that atom cannot be excited.  This is the basic idea behind the Rydberg blockade concept of \citet{Jaksch2000}.

\begin{figure}
\includegraphics[width=2.5 in]{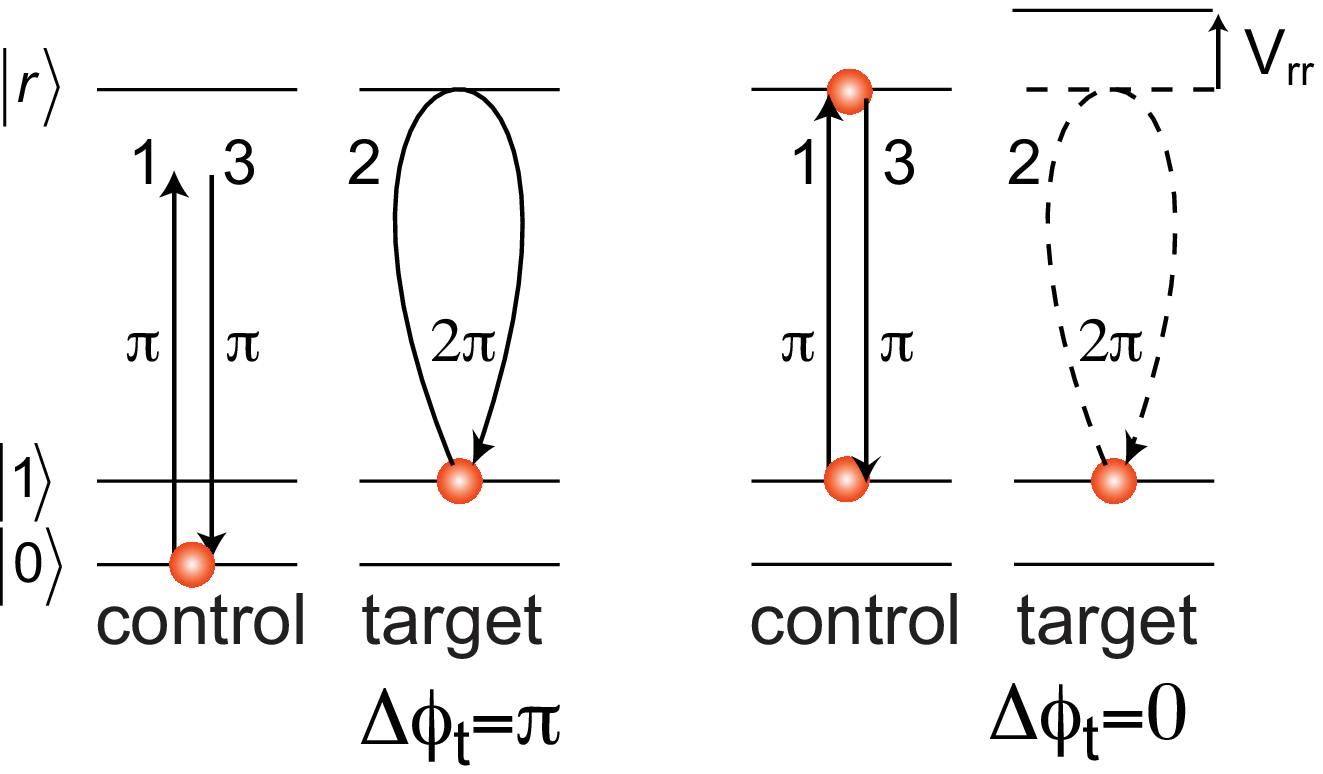}\hspace{0.2 in}
\includegraphics[width=2.5 in]{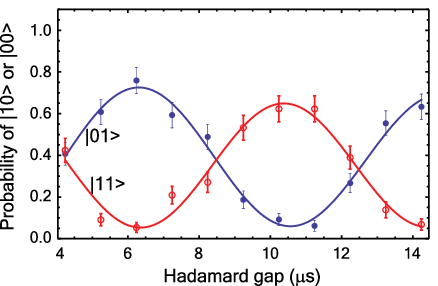}
\caption{a) Protocol for controlled-phase gate.  b) Experimental demonstration of the $\pi$ phase shift induced in the target atom by changing the state of the control atom. From \protect\cite{Isenhower2010}.}\label{cphase}
\end{figure}

The basic entangling operation under Rydberg blockade conditions is the controlled-phase gate.  The protocol is shown in Fig.~\ref{cphase}.  First, the ``control'' atom is excited via a $\pi$ pulse from $\ket{1}$ to $\ket{r}$.  In our experiments this is done by an off-resonant two-photon excitation, as will be explained in Section~\ref{sec:RydRab}.  According to the standard Rabi evolution, the $\ket{1}$ component of its wave function is changed to $-i\ket{r}$.  Then the ``target'' atom is subject to a 2$\pi$ Rydberg excitation pulse. Its $\ket{1}$ component is thereby changed to $-\ket{1}$, unless the control atom was in the Rydberg state, in which case its $\ket{1}$ component is nominally unchanged.  Finally, a second $\pi$ pulse is applied to the control atom to return it from the Rydberg state to state $-\ket{1}$.  Thus, in the ideal case the controlled phase gate implements the mapping
\be
C_Z={\rm diag}(1,-1,-1,-1)
\ee
on the two-atom basis states $\ket{00}$, $\ket{01}$,  $\ket{10}$, $\ket{11}$.  Experimental demonstration of the controlled phase shift from \citet{Isenhower2010} is shown in Fig.~\ref{cphase}.

In reality, since the blockade interaction is not infinite, there are important corrections to the performance of the gate.  For the $\ket{11}$ input, the finite blockade results in an AC-Stark shift { $\Omega^2/(4V_{rr})$ of the $\ket{r1}$ level}, where $\Omega$ is the two-photon Rabi frequency.  This results in a phase shift $\phi_{rr}=\Omega^2/(4V_{rr})(2\pi/\Omega)=\pi\Omega/2V_{rr}$
of the $\ket{11}$ entry.  This can in principle be compensated for \citep{Zhang2012} by changing the 2$\pi$ target pulse to a $\pi$-delay-$\pi$ pulse, where the delay is chosen in concert with the normal single-atom AC Stark shifts to give an equal phase $\phi_{rr}$ to the $\ket{01}$ input.  This gives an overall phase shift $\phi_{rr}$ of the target atom $\ket{1}$ state, which can be removed by a single-atom rotation.

A more serious error in the phase gate occurs as the finite blockade strength allows a small population to accumulate in the $\ket{rr}$ state via off-resonant excitation.  The probability of double excitation
 is about $\pi\Omega^2/2V_{rr}^2$. In principle, this can be reduced by going to higher principle quantum number to increase $V_{rr}$ or by reducing the two-photon Rabi frequency.  These choices must be weighed against any deleterious effects produced by increased spontaneous emission.  We have recently given a detailed analysis of these tradeoffs in \citet{Zhang2012}.

The controlled-phase gate is itself an entangling gate.  When sandwiched between two Hadamard gates {(single-atom $\pi/2$ rotations)}, it becomes a C-NOT gate, a widely used fundamental gate for quantum computation.

At this point, it is useful to summarize the attractive features of {the Rydberg blockade mechanism for entangling} atoms at optically resolvable distances.
Table~\ref{tab:cphase}  traces the evolution of the various two-atom input states through the phase gate protocol.   Notice that at no point in the protocol are both atoms in the Rydberg state.  Thus no real interatomic forces are experienced by the atoms.  The entanglement arises not from a state-dependent controlled interaction between the two atoms, but by a (large) frequency shift of a thereby virtually accessed state.

Note also that the strength of the Rydberg-Rydberg interaction does not have to be precisely controlled.  The errors due to variations in the blockade strength are small, provided the blockade interaction greatly exceeds the two-photon Rabi frequency.

\begin{table}
\begin{center}\begin{tabular}{|c|c|c|c|}\hline
Input&$\pi$-control&$2\pi$-target&$\pi$-control \\ \hline
$\ket{00}$&$\ket{00}$&$\ket{00}$&$\ket{00}$ \\ \hline
$\ket{01}$&$\ket{01}$&-$\ket{01}$&-$\ket{01}$ \\ \hline
$\ket{10}$&-i$\ket{r0}$&-i$\ket{r0}$&-$\ket{10}$ \\ \hline
$\ket{11}$&-i$\ket{r1}$&-i$\ket{r1}$&-$\ket{11}$ \\ \hline
\end{tabular}\caption{Evolution of 2-atom input states in the controlled phase blockade protocol.}\label{tab:cphase}
\end{center}\end{table}

The blockade mechanism produces entanglement between the spin degrees of freedom of the atoms.  It is therefore important to avoid unintentional coupling to external degrees of freedom by having state-dependent forces (see Sec.~\ref{sec.Rydbergtrap}).  In particular, the polarizability of Rydberg states is negative, while the polarizability of the qubit states is positive for the types of FORTs used here.  To avoid applying state-dependent forces, we turn off the FORTs during the blockade gate.  

\section{Trapping and Readout of Single Atoms}\label{sec:trapping}
\begin{figure*}[!t]
\centering
\includegraphics[width=14.cm]{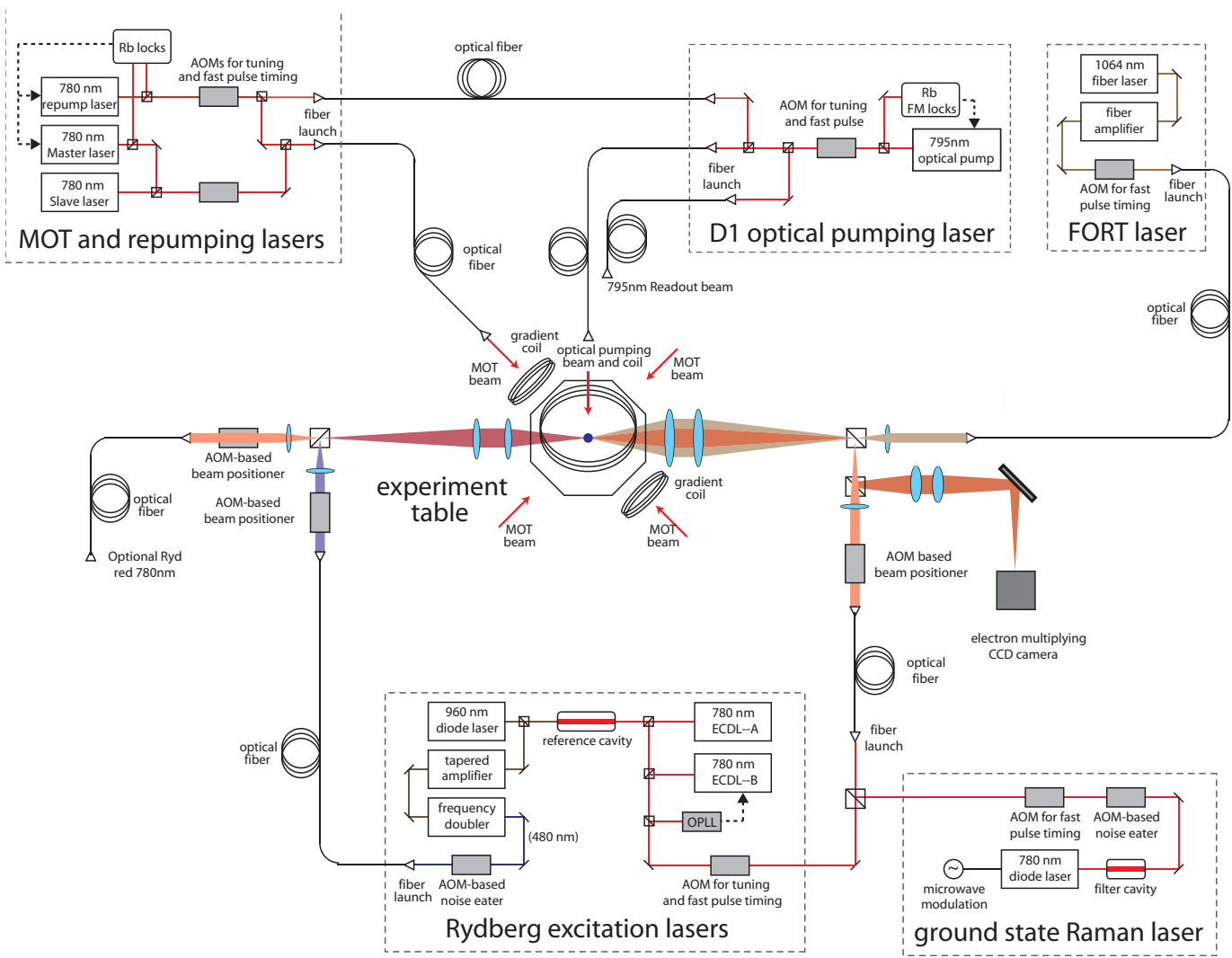}
\caption{Experimental setup for Rydberg gate experiments.
}
\label{fig.setup}
\end{figure*}

Quantum gate and entanglement experiments with single atoms have required a fairly complex experimental apparatus. Multiple lasers covering a large range of wavelengths are focused to $\mu\rm m$ sized spots that must be aligned relative to each other with high precision for atom trapping, readout, and control. Fast spatial and temporal modulation of the lasers is needed, together with good control of the amplitude and frequency characteristics in order to achieve high fidelity operations.  An overview of the experimental apparatus is shown in Fig. \ref{fig.setup}. We will refer to this figure in several of the subsequent sections. While a variety of neutral atoms have been successfully laser cooled and trapped, entanglement and qubit experiments are most developed for the heavy alkalis Rb and Cs, {whose large fine-structure splittings are advantageous for state preparation and readout}. We will therefore give orders of magnitude and characteristic quantities
 in the following discussions that are relevant to experiments with heavy alkali atoms.

\subsection{Optical traps}

Experiments with neutral atom qubits   build on
well developed  techniques for cooling and trapping atoms with electromagnetic fields \citep{Metcalf1999,Foot}. Using standard laser cooling techniques it is straightforward to create small clouds of cold alkali atoms in a magneto-optical trap (MOT) with temperatures below $50~\mu\rm K$ and densities in the range of $10^9 - 10^{11}~\rm cm^{-3}$. Single atoms can then be transferred from the MOT into a FORT for qubit experiments. The most straightforward way of doing this is simply to overlap a red detuned, tightly focused beam with the MOT cloud, and then turn off the MOT cooling lasers and inhomogeneous magnetic field. A useful overview of  optical traps for atoms can be found in \cite{Grimm2000}.

For ground state alkali atoms the trapping potential created by the FORT is given 
by $U_F=-\frac{1}{4}\alpha |{\mathcal E}|^2$, where $\alpha$ is the polarizability and ${\mathcal E}$ is the electric field amplitude.  The polarizability should be calculated by accounting for all allowed transitions to excited atomic states. In the heavy alkalis the polarizability is dominated by the contributions from the first resonance lines and a useful approximation for large detuning is 
\be
U_F({\bf r})=\frac{\pi c^2 \Gamma}{2\omega_a^3}\left(\frac{2}{\Delta_{3/2}}+\frac{1}{\Delta_{1/2}} \right)I(\bfr).
\ee
Here $\omega_a$ {and $\Gamma$ are the frequency and decay rate }of the  $5S_{1/2}-5P_{3/2}$ transition, $\Delta_{3/2(1/2)}$ are the laser detunings from the $5 P_{3/2(1/2)}$ states and $I(\bfr)$ is the trapping intensity.
The peak intensity in a Gaussian beam with power $P$ and waist $w$ is $I(0)=2P/\pi w^2$. For Rb we find, for example,  that a trap depth of
1 mK requires $48~\rm  mW$ of power in a 1064 nm laser beam focused to $w=2~\mu\rm m$.

The advantage of loading single atoms into optical traps is that they are localized to a region with dimensions that scale as 
$\delta x=\delta y\sim w \sqrt{k_B T/U_F}$ transversely and  $\delta z\sim (w^2/\lambda_F) \sqrt{k_B T/U_F}$ axially along the propagation direction of the FORT beam.
 Explicit expressions for the localization lengths and oscillation frequencies can be found in \cite{Saffman2005a}. With $T\sim 50~\mu\rm K$ and $U_F\sim k_B\times 1 ~\rm mK$ it is possible to obtain sub-micron transverse localization and few micron axial localization in this type of simple single beam trap which serves as an excellent starting point for qubit experiments. 

Optical traps are attractive, not only for their localization properties, but also because decoherence rates can be very low. 
Photon scattering rates scale as $1/\Delta^2$ and can be much less than $0.1~\rm s^{-1}$ for typical parameters as given above. 
At large detuning interaction with the trap light leads mostly to Rayleigh scattering which does not {appreciably }decohere the qubit state, but does give motional heating. Raman processes which lead to transitions $\ket{0}\leftrightarrow \ket{1}$ as well as transitions to Zeeman states outside of the computational basis scale as (\cite{Cline1994}) $1/\Delta^4$, and are thus even further suppressed compared to Rayleigh scattering.

{One complication of the intense FORT laser is substantial AC Stark shifts of the excited-state as well as ground-state energies.  The resonance positions can by altered by many line widths, causing problems with cooling.  To mitigate this we alternate the FORT lasers with the cooling and readout lasers, thus only scattering light when the FORT is off. }

\subsection{Detection of single atoms and quantum states}

\begin{figure}[!t]
\includegraphics[width=6cm]{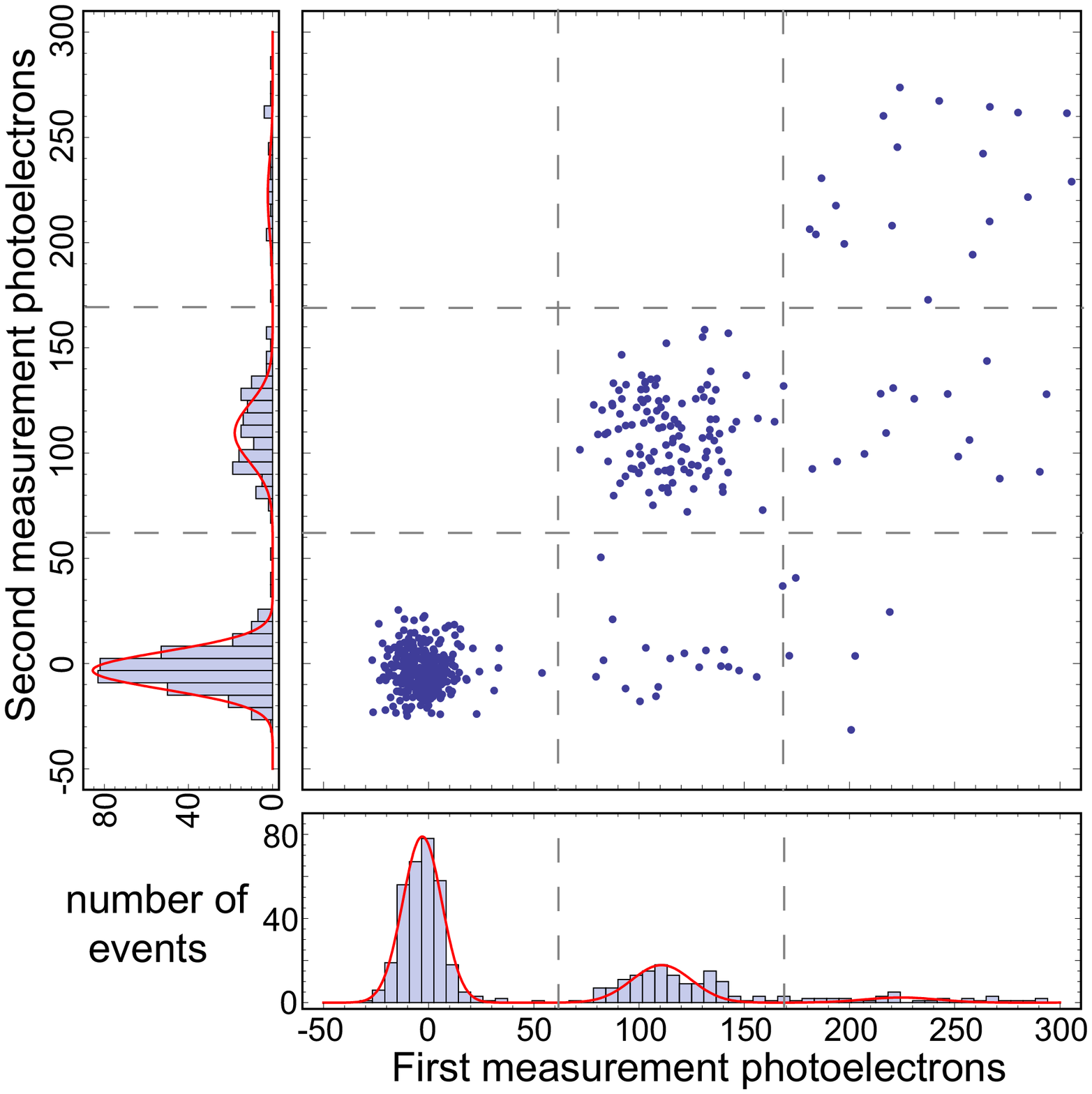}
\includegraphics[width=8.cm]{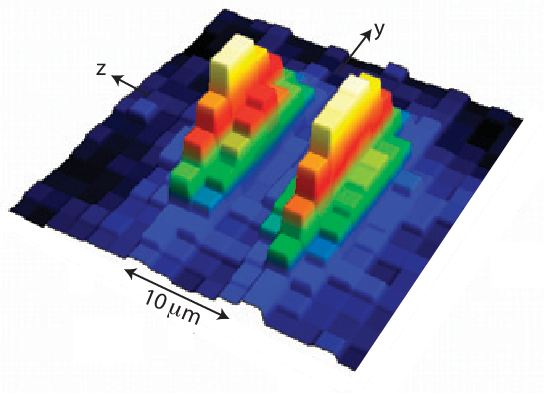}

\caption{{(left) Histogram of photon counts from trapped atoms from \cite{Johnson2008}, showing also the ability to clearly distinguish a single atom from the background and to detect the presence of one atom without removing it from the trap.}   {(right) Multiple exposure photograph of the spatial distribution of single atoms in the two FORT sites, from \protect\cite{Urban2009}.}
}
\label{fig.atomdetection}
\end{figure}

Detection of single trapped atoms can be performed by detection of resonance fluorescence scattered from the atom. While other methods such as single atom absorption or phase shifts are also feasible (\cite{Wineland1987, Aljunid2009}), resonance fluorescence is most widely used and is the most straightforward to implement. The maximum scattering rate when the transition is strongly saturated is $\Gamma/2$, and typical overall photon detection efficiencies accounting for collection lens solid angle, optical transmission losses, and detector quantum efficiency are about $1\%$. In order to make a reliable measurement it is necessary to detect enough photons such that the count fluctuations due to Poissonian statistics, background scattering from optical elements in the setup, and detector dark counts are much smaller than the counts due to the presence of an atom. For alkali atoms the above estimates give detector count rates  $\sim 20~\rm ms^{-1}$, so measurement times are generally a few  ms. A count rate histogram showing clear separation between zero and one atom signals is shown in Fig. \ref{fig.atomdetection}. The large separation between the zero and one atom peaks implies a readout fidelity well above  95\%.

The maximum scattering rate for a single atom is $\Gamma/2$ if the transition is strongly saturated. Using three dimensional molasses beams tuned to the red of the scattering resonance  the atom is cooled while scattering photons, which tends to prevent atom loss during readout. Velocity fluctuations that naturally occur during laser cooling can nevertheless lead to atom loss. We have therefore used trap depths that are  1 mK, or larger, to limit atom loss during readout.  {Figure~\ref{fig.atomdetection} demonstrates low-loss atom readout, and shows a composite image of the spatially resolved atom distribution.}

Transfer of single atoms from a MOT to a FORT by spatial overlapping is stochastic in nature with the number distribution  governed by Poissonian statistics. One would therefore expect that the maximum possible success rate for loading a single atom 
is $1/e\simeq 0.37$. It turns out that in the regime of very small FORTs with waists under $1~\mu\rm m$, strong collisional interactions tend to remove pairs of atoms, { leading to sub-Poissonian atom number statistics} (\cite{Schlosser2001}) and  a higher probability of singles with very few two-atom events. Single atom loading probabilities of about 50\% can be observed over a large range of loading rates (\cite{Schlosser2002}).  { While our more weakly confining FORTs do typically give some moderately sub-Poissonian loading,  we generally run these experiments with about 30\% single atom loading probability. For two-qubit experiments this limits the useful data collection to a duty cycle of about 10\%.}

\subsection{{Single-atom state detection}\label{sec:state}}

Detection of the quantum state of a single atom is more challenging due to the fact that unwanted Raman transitions occur which transfer an atom cycling on the 
{nominally closed $5S_{1/2}(F=2) \rightarrow 5P_{3/2}(F=3)$ transition to $n_gS_{1/2}(F=1)$. }
The rate of Raman transfers $r_R$ relative to the rate of cycling transitions $r_c$ scales as $r_R/r_c\sim (\Gamma/\Delta_{\rm hf})^2$. Detailed calculations that account for branching ratio factors show $r_c/r_R\sim 40,000$ for $^{87}$Rb. 
The probability of detecting $N$ photons without suffering a Raman event using a system with 1\% detection efficiency  is thus $P_{N}= e^{-100 N r_R/r_c}\simeq 0.78$ {for N=100} which implies only moderately good state detection fidelity.  Discriminating between quantum states with high fidelity requires high collection efficiency and low backgrounds, so that the number of scattered photons can be reduced.  This challenge was met in two recent experiments which demonstrated single atom state detection with better than 95\% efficiency and only 1\% atom loss (\cite{Gibbons2011,Fuhrmanek2011}).

{As our} detection efficiency is not sufficiently high to enable state detection without atom loss, {we} distinguish $\ket{0}$ and $\ket{1}$ qubit states using a lossy method (\cite{Kuhr2003}).
The idea is to push atoms out of the trap in a state selective fashion using a beam that is resonant with  the 
{$5S_{1/2}(F=2) - 5P_{3/2}(F=3)$} cycling transition. The number of scattered photons needed for the mechanical push is much less than the number needed for state detection so this can be done with high selectivity between states despite the Raman rate mentioned above. While this is effective and has been used as the method of choice in the Rydberg quantum gate experiments to date (\cite{Isenhower2010,Wilk2010,Zhang2010}) it is not a quantum nondemolition (QND) measurement. The practical drawback is that a new atom must be loaded after each measurement, which greatly reduces the obtainable data rate.

\subsection{Optical trap {effects on} Rydberg atoms}

\label{sec.Rydbergtrap}

Rydberg blockade experiments  rely on transient excitation to Rydberg states to turn on strong interatomic potentials. Although the atoms are only Rydberg excited for a few $\mu\rm s$ to avoid decoherence due to spontaneous emission, it is still necessary to account for the trap induced  potential seen by a Rydberg atom. Highly excited Rydberg states 
may be approximated  as a quasi-free electron which has a negative ponderomotive polarizability $\alpha_e= -e^2/m_e \omega^2$ \citep{Dutta2000}, with $\omega$ the frequency of the trapping light. Red detuned FORTs rely on a positive atomic ground state polarizability so the Rydberg states are repelled by the FORTs. Even if the time spent in a Rydberg excited state is too short to mechanically eject the atom from the trap the difference in polarizabilities leads to a position dependent excitation frequency which broadens the ground to Rydberg transition, and impacts the fidelity of quantum state transfer. Furthermore, the trap light photoionizes Rydberg states. {For these reasons our experiments to date have turned off the FORT light during Rydberg manipulations}. Provided the atoms are sufficiently cold they can be recaptured by turning the trap back on again immediately following the Rydberg pulses, after the atom has been returned to the ground state.   
 
Despite the fact that the atom can be recaptured with high probability this situation is far from ideal for quantum logic experiments since the heating that arises from transfer between different external potentials leads to entanglement between  spin and motional degrees of freedom. 
This can be seen by the following simple argument. Suppose the qubit state
$\ket{\psi}=a\ket{0}+b\ket{1}$ is stored in an atom in the ground state $\ket{0}_{\rm vib}$ of the trapping potential. The total state of the qubit plus atom
is $\ket{\Psi}=\ket{\psi}\otimes\ket{0}_{\rm vib}.$
Vibrational excitation during a Rydberg cycle  will lead to the new state
$\ket{\Psi'}=a\ket{0} \otimes \ket{0}_{\rm vib}+b\ket{1} \otimes ( c\ket{0}_{\rm vib}+d\ket{1}_{\rm vib})$ where, for simplicity,
we have only considered excitation of the first vibrational state with amplitude $d.$
Tracing over the vibrational degrees of freedom gives the reduced density matrix
\be
\rho_{\rm qubit}={\rm Tr}_{\rm vib}[\rho]=
\begin{pmatrix}
|a|^2& ab^* c^*\\
a^*b c&|b|^2
\end{pmatrix}.
\label{eq.qubitvib}
\ee
Since $|c|< 1$ vibrational excitation results in reduced coherence of the qubit.
As discussed in Sec.~\ref{sec:concl}, one anticipated advantage of dark FORTs is a substantial reduction of this effect.

\section{State Preparation}\label{sec:stateprep}

Among the $2(2I+1)$ magnetic sublevels of the alkali ground state, the two states with $m_F=0$ are  obvious choices for qubit states due to their weak quadratic sensitivity to magnetic field fluctuations.  To second order in the magnetic field $B_z$, the energies of these states are
\be
E(F)= \left\{  
\begin{array}{cc}
 h\nu+\frac{\mu_B^2B_z^2}{h\nu }  &F= I+\frac{1}{2} \\
 -\frac{\mu_B^2B_z^2}{h\nu } & F=I-\frac{1}{2}
\end{array}
\right.
\ee
where $\nu=6834.68$ MHz is the zero-field clock frequency for $^{87}$Rb, and $\mu_B=1.4$ MHz/G is the Bohr magneton.  Here and elsewhere we express energies in frequency units.  When discussing these two states as qubits, we refer to the state $\ket{S_{1/2}(F=I+1/2,m_F=0)}\equiv\ket{1}$ and $\ket{S_{1/2}(I-1/2,0)}\equiv\ket{0}$. In our laboratory, with no magnetic shielding, measurements of $T_2$ in Ramsey experiments at different bias fields suggest magnetic field fluctuations at the 25-50 mG level, implying a potential $\sim 1$ Hz line width for the 0--0 resonance, assuming no other decoherence sources \citep{Saffman2010}.  Thus this qubit choice has the potential for extremely high performance (see Sec.~\ref{sec:FORTdecoh} for a further discussion).

A key component of any coherent quantum protocol is proper initialization of the qubits.  This involves first putting the atoms into one of the qubit states using optical pumping, and then making arbitrary superpositions of the two qubit states using stimulated Raman transitions.  Raman transitions are also used by many gate protocols, so they are invaluable for computations themselves.

\subsection{Optical Pumping}

To initialize a computation, we need to optically pump the qubit into one of the qubit states.  The simplest way to do this is to apply $\hat{z}$-polarized light, { parallel to a $\sim3$ Gauss magnetic field}, on the $S_{1/2}(F=2)\rightarrow P_{1/2}(F=2)$ transition. Repumper light (typically $S_{1/2}(F=1)\rightarrow P_{3/2}(F=2)$) is also necessary to eliminate $F=1$ populations.  Due to the zero matrix element $\langle P_{1/2}(2,0)|z|S_{1/2}(2,0)\rangle$  the atoms tend to accumulate in the 
$|S_{1/2}(2,0)\rangle$ state by optical pumping.  

We model the optical pumping process by assuming that there is an optical pumping rate $R_{\rm op}$ into the $\ket{1}$ state, and a depumping rate $R_1$ out of the $\ket{1}$ state.  The optical pumping transients then obey
\be
{dP_1\over dt}=R_{\rm op}(1-P_1)-R_1P_1
\ee
with solution
\be
P_1(t)={R_{\rm op}\over R_{\rm op}+R_1}\left(1-e^{-(R_{\rm op}+R_1)t}\right)+P_1(0)e^{-(R_{\rm op}+R_1)t}
\ee

Simulations \citep{HJW} show that the mean number of photons required to pump initially unpolarized atoms into the $\ket{1}$ state is about 4.5, so $R_{\rm op}\approx R/4.5$, where $R$ is the scattering rate for unpolarized atoms.

\begin{figure}
\includegraphics[width=5 in]{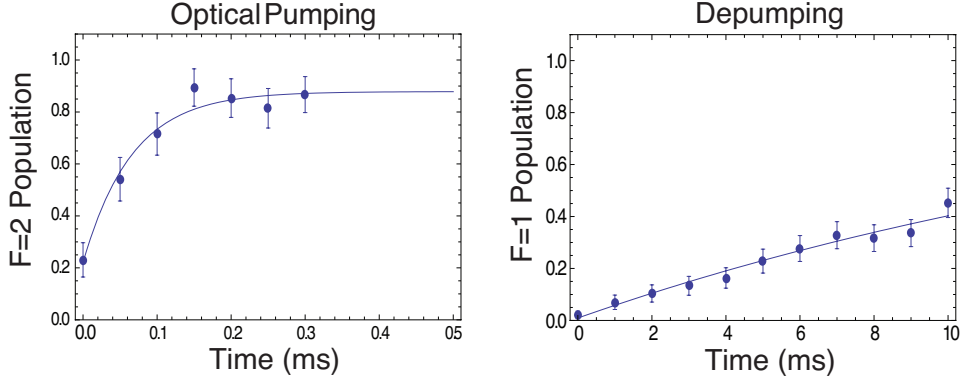}
\caption{Optical pumping transients for preparation of the $\ket1$ qubit level. Left: with both pumping lasers on, the atoms rapidly accumulate in the $F=2,m=0$ ($\ket1$) level.  Right: blocking the $F=1$ laser allows measurement of the rate at which atoms are removed from the $\ket1$ state.  }\label{fig:pumping data}
\end{figure}

A simple method for estimating the optical pumping fidelity is to measure the pumping time constant $\tau=1/(R_{\rm op}+R_1)$ and the depumping rate $R_1$ deduced by measuring the decay of population out of state $\ket{1}$ with $R_{\rm op}=0$ (this is done by turning off the laser that depletes the $F=1$ states).  Then
\be
P_1(\infty)=1-R_1\tau.
\ee
Typical data for this method are shown in Fig.~\ref{fig:pumping data}.

There are two effects that fundamentally limit the quality of the optical pumping.  The finite excited-state hyperfine splitting results in $R_1>0$ (because of Raman scattering out of the $\ket{1}$ state via the $\ket{P_{1/2}(1,0)}$ level).  A second effect is that the magnetic field mixes a small amount of the $\ket{P_{1/2}(1,0)}$ state into the $\ket{P_{1/2}(2,0)}$ state, making the excitation rate to the $\ket{P_{1/2}(2,0)}$ state non-zero.  Other technical issues, such as misalignment of the pumping beam polarization with respect to the magnetic field and imperfect linear polarization of the pumping beam, can also limit the fidelity.  It should be possible to achieve better than 99.9\% pumping of the atoms into the $\ket{1}$ state if these technical problems are carefully addressed.

\subsection{Single qubit rotations}

\begin{figure}
\includegraphics[width=2.7 in]{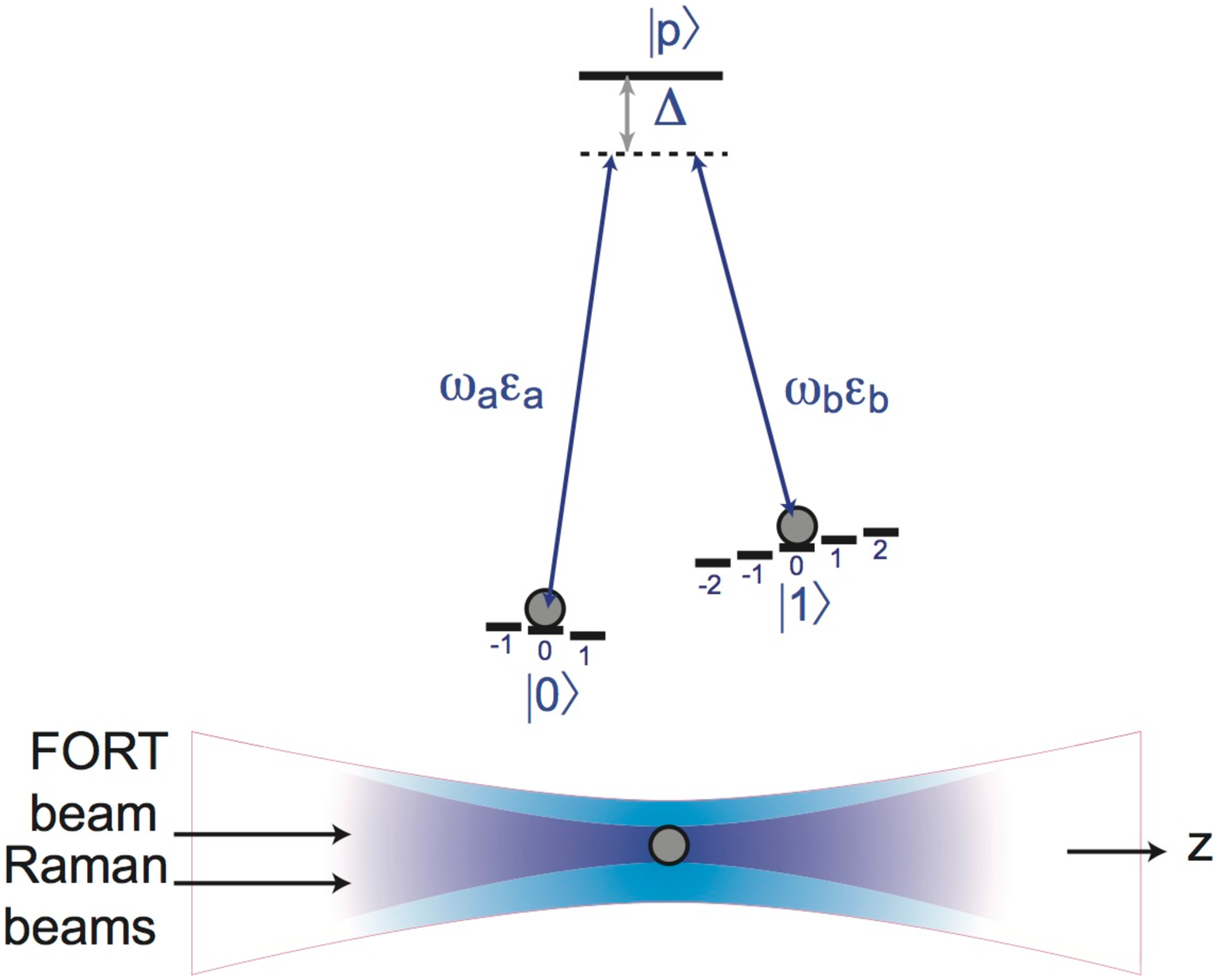}
\includegraphics[width=2.7 in]{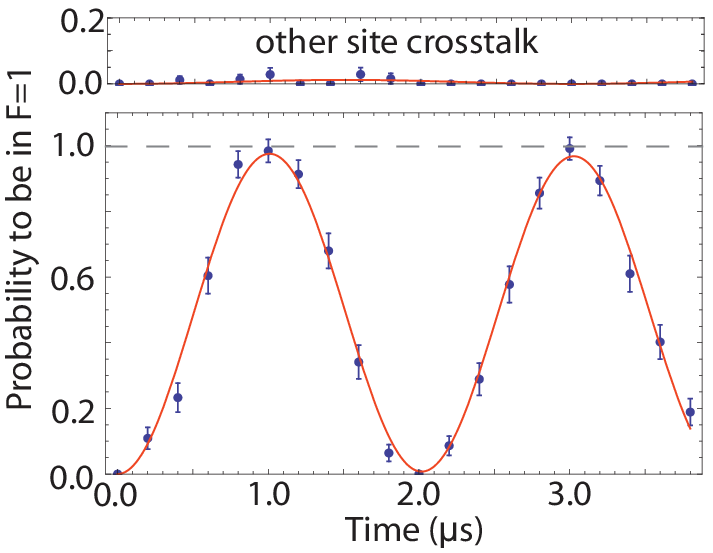}
\caption{Addressable single-atom rotations using stimulated Raman scattering, adapted from \cite{Saffman2005a}.  Left: energy levels and laser geometry.  Right:  Rabi flopping data, including cross-talk.  Adapted from \protect\cite{Zhang2010}.}\label{fig:Raman}
\end{figure}

Stimulated Raman transitions between the two qubit states are driven by a pair of $\sigma^+$ polarized light waves whose frequency difference is very nearly equal to the free qubit Bohr frequency, see Fig.~\ref{fig:Raman}.  In order for the transitions to be highly coherent, it is desirable to detune the light quite far from the excited-state resonance--roughly $\Delta\sim 2\pi
\times100$ GHz for these experiments.  Under such conditions, the amplitude of the excited states is very small, roughly $\Omega/\Delta$, and the excited state amplitudes adiabatically follow the ground state amplitudes.  In this limit, the system behaves very nearly as a two-level system with an effective Rabi frequency
\be
\Omega_1={\Omega_a\Omega_b\over 2 \Delta}
\ee
There are two primary complications to this simple picture: AC Stark shifts and spontaneous emission.

Spontaneous emission out of the excited level gives the two qubit levels finite decay rates $\delta R_i\approx \Gamma\Omega_i^2/4\Delta^2$.  This has two effects.  First, the coherent superpositions produced by the Raman rotations decay with an effective time $T_2\approx 2/(\delta R_0+\delta R_1)$.  Second, the scattering removes population from the qubit basis into other hyperfine levels.  The probability of this happening in a $\pi$-pulse is approximately $\delta R\pi/\Omega_1=\pi\Gamma/2\Delta$, showing the importance of a large detuning for high fidelity single atom rotations.

The qubit levels are also AC Stark shifted by an amount $\delta U_i=\Omega_i^2/2\Delta$, giving an AC Stark shift of the qubit frequency of $\delta\omega=(\Omega_1^2-\Omega_1^2)/2\Delta$, which cancels when the individual Rabi frequencies of the two lasers are equal.

There are two advantages to only partially canceling the AC Stark shift.  If the Raman beams are tuned to be resonant with the AC Stark-shifted qubit resonance, this implies that when the Raman beams are not applied, the qubit precesses with respect to the microwave phase at a rate $\delta\omega$.  This is a very simple method for implementing qubit rotations about the z-axis.
The AC Stark shift also helps to suppress cross-talk between the two qubits.  

{We generate the Raman light by microwave modulation of the drive current of a diode laser at 1/2 the qubit resonance frequency.  The laser carrier frequency is tuned typically 90 GHz above the $P_{3/2}$ resonances.  The laser is then coupled through a moderate finesse cavity to strip the carrier and high order sidebands, leaving just the the $\pm1$ order sidebands needed to drive the qubit resonance.  A key feature of our apparatus is the ability to selectively address the individual qubits using dynamically frequency-shifted acousto-optic modulators.}

Figure~\ref{fig:Raman} shows experimental measurements of the ground-state flopping, including the very small cross-talk observed.  The main improvements over our original demonstration of this capability \citep{Yavuz2006} involve improved optical pumping and readout.  At present, the quality of this part of the process is sufficient to be considered a negligible contributor to experimental errors.

\section{Coherent Rydberg Rabi Flopping}\label{sec:RydRab}

Coherent transfer of quantum states between ground and Rydberg levels can be performed by laser excitation. The energy difference exceeds 1 petaHz
and the wavelength needed for single photon excitation to $n=100$ is $297  ~\rm nm$ in Rb. Building high power, narrow linewidth sources at this short wavelength is challenging, so we, and many other research groups have relied on two-photon excitation to $nS$ or $nD$ states. The experimental approach we have followed with Rb atoms is shown in Fig. \ref{fig.setup}. A diode laser at 780 nm is
combined with the second harmonic of a 958 nm diode laser giving a two-color source to drive the $5S_{1/2} - 5P_{3/2}$  transition at 780 nm, followed by $5P_{3/2} - nS_{1/2}, nD_{3/2},$ or $nD_{5/2}$ using 479 nm. In this section we focus on issues relevant for single atom quantum logic experiments. A more extensive discussion of this topic can be found in \cite{Saffman2010}.

 When the intermediate level detuning $\Delta_p=\omega_1-\omega_{ps}$ is large compared to the width of the hyperfine structure of the
$\ket{p}$ level
the two-photon Rabi frequency  is given by $\Omega=\Omega_{1} \Omega_{2}/2\Delta_p$. The one-photon  Rabi frequencies are $\Omega_1=-e{\mathcal E}_1\bra{p} {\bf r}\cdot {\bfepsilon}_1\ket{s}/\hbar$,
$\Omega_2=-e{\mathcal E}_2\bra{r}{\bf r}\cdot {\bfepsilon}_2\ket{p}/\hbar$,
with ${\mathcal E}_j, \bfepsilon_j$ the field amplitudes and polarizations.  The transition matrix elements
can be reduced via the Wigner-Eckart theorem to  an angular factor times a radial integral $\langle r \rangle$. 
For the $s-p$ transition the radial integral is known (for $^{87}$Rb $\langle r\rangle_{5s_{1/2}}^{5p_{3/2}}=5.18\times  a_0$) and for the $p-r$ transition it can be calculated numerically (\cite{Walker2008}).  The following  expressions are accurate to better than 10\%
for $^{87}$Rb and $n>50$:  $\langle r\rangle_{5p}^{ns}=.014 \times (50/n)^{3/2}a_0$ and $\langle r\rangle_{5p}^{nd}=-.024 \times (50/n)^{3/2} a_0$.

Population of the intermediate $\ket{p}$ level during two-photon excitation results in spontaneous emission and loss of coherence. The probability of this occurring during a $\pi$ excitation pulse of duration $t=\pi/|\Omega|$ is
$P_{\rm se}=\frac{\pi\gamma_p}{4|\Delta_p|}\left(q+\frac{1}{q}\right)$
where $q=|\Omega_2/\Omega_1|.$ The spontaneous emission is minimized for $q=1$ which lets us write the
Rabi frequency as
\be
\Omega=\frac{P_{\rm se}}{\pi}\frac{|\Omega_2|^2}{\gamma_p}.
\label{eq.2photonRabi}
\ee
We see that fast excitation with low spontaneous emission is possible provided $\Omega_2$ is sufficiently large. This is increasingly difficult as $n$ is raised since $\langle r \rangle_{5p}^{nl}\sim 1/n^{3/2}$.
Put another way, at constant $\Omega$ and $P_{\rm se}$ the required optical power scales as $n^3$.

Experimental requirements on laser stability and linewidth can be estimated as follows. The error probability when transferring population from  ground to Rydberg states using a $\pi$ pulse is proportional to $\Delta^2/\Omega^2$, where $\Delta$ is the two-photon detuning. With $\Omega/2\pi \sim 1~\rm MHz$ an excitation error of $10^{-4}$ requires $\Delta/2\pi \stackrel{<}{\sim}10~\rm kHz$. The lasers used should therefore be long term frequency stabilized to under 10 kHz in order to acquire data over several hours, and also have short term linewidths under 10 kHz for the few $\mu\rm s$ timescales of the Rydberg pulses. These requirements can be met using very high finesse optical cavities constructed from ultra-low expansion (ULE) glass  as frequency references. As can be seen in Fig. \ref{fig.setup} we have used a single ULE cavity with mirror coatings 
at both 780 and 960 nm. The cavity finesse was $>10^5$ at both wavelengths. The cavity is placed in an ultra high vacuum can which is temperature stabilized to a few mK. This system gives short term linewidths well under 1 kHz, and drifts over a few hours at the level of 
$\sim 10 ~\rm kHz$. Improved long term stability could be obtained by operating the cavity at the zero thermal expansion temperature (which may differ from the nominal 25 C of ULE) and improving the thermal control as in \cite{Alnis2008}.

In addition to laser stabilization the atomic transition frequency must also be stable. This requires careful attention to fluctuations of external electric and magnetic fields, laser induced AC Stark shifts, and Doppler broadening. To a good approximation the AC Stark shifts are dominated by the 780 nm laser interacting with the ground state and the 
479 nm laser interacting with the Rydberg state. These shifts are proportional to $\delta U_g\sim \Omega_{1}^2/\Delta_p$ and $\delta U_r\sim \Omega_{2}^2/\Delta_p$. Provided the one-photon Rabi frequencies $\Omega_1, \Omega_2$ are equal, the AC Stark shifts are also equal, and there is no shift of the transition frequency. 
This is technically challenging as the $p$ -  Rydberg matrix element is much smaller than that of the ground - $ p$ excitation. In the absence of Stark shift cancellation it is necessary to have well stabilized 
laser intensities, and to limit motion of the atoms under the envelope of the exciting laser beams, which implies the atoms should be cold and well localized. We have used acousto-optic modulator based ``noise eater" devices to reduce intensity noise on all lasers used for state control.  

{Electric field fluctuations are also potentially a serious issue in Rydberg experiments near $n=100$.  We have not yet seen evidence for significant stray fields inside our grounded stainless steel vacuum chamber.}

Doppler broadening is reduced by using a counterpropagating arrangement for the beams. In this way 
$k_2 = k_{479}-k_{780}$ is reduced by a factor of 4.2 compared to $k_{479}+k_{780}$ which would apply for copropagating beams. Since Doppler broadening scales as $k_2 v\sim k_2 T^{1/2}$ using counterpropagating beams effectively reduces the kinetic 
temperature contributing to broadening by a factor larger than 16.
It was recently proposed (\cite{Ryabtsev2011}) to use three photon excitation of $nP$ states in a star like configuration to simultaneously eliminate Doppler broadening and photon recoil shifts. Despite the added complexity this idea may prove particularly valuable for  coherent Rydberg excitation in hot atomic samples (\cite{Huber2011}).

In order to get a sense of experimentally relevant errors consider the following example of excitation of
the $^{87}$Rb $100d_{5/2}$ level via $5p_{3/2}$. Let us assume $\pi$ polarized beams with powers of  $1~\mu\rm W$ at 780 nm and $300~\rm mW$ at 480 nm
 focused to spots with Gaussian waist $w=3~\mu\rm m$.
This gives single photon Rabi frequencies of $225\, \&\, 210~\rm MHz.$
The light is detuned from $5p_{3/2}$ by $\Delta_p/2\pi = 20~\rm GHz$. These parameters couple  $m=0$ ground states to $m=\pm 1/2$ Rydberg states with a Rabi frequency
$\Omega/2\pi = 1.2~\rm MHz$. The probability of spontaneous emission from the $p$ level during a $\pi$ pulse
is $P_{\rm se}=5\times 10^{-4}. $
 The fractional excitation error after a $\pi$ pulse due to Doppler broadening is $P_{\rm Doppler}=|\delta/\Omega|^2 $. For $^{87}$Rb atoms at $T=10~\mu\rm K$ and counterpropagating   excitation beams we find
$P_{\rm Doppler}=4\times 10^{-4} $. Thus, coherent excitation of a  high lying Rydberg level   with
combined spontaneous emission and Doppler errors
below $10^{-3}$ is within reach of current experimental capabilities.

Nevertheless a challenging experimental aspect is the need {to deliver} rather high power (300 mW)  narrow band 480 nm light at the atoms. Although this power level can be produced routinely, temporal and spatial modulation of the laser beam typically introduces large attenuation factors. The system in Fig. \ref{fig.setup} easily generates 200 mW of single frequency 479 nm radiation, but after noise eaters, pulse control, fiber transfer, and spatial switching only 15-20 mW are available at the atoms.  {As discussed in Sec.~\ref{sec:concl}, this situation should be greatly improved by excitation via the 2nd resonance lines}

In addition to the above  it is necessary to consider the laser polarization and choice of Rydberg Zeeman states. We are coupling hyperfine ground states
characterized by quantum numbers $n,I,j,l,s,f,m_I,m_f$ to highly excited Rydberg fine structure states that have negligible hyperfine structure, and are therefore described by the quantum numbers $n',I,j',l',s,m_I',m_j'.$
In most cases there are two Rydberg Zeeman states with different values of $m_j'$ that have nonzero electric dipole matrix elements with the ground state. Only one Rydberg Zeeman state is coupled to if we start from a stretched ground state $m_f=\pm f$ or use $\omega_1$ with $\sigma_\pm$ polarization coupling via a $np_{1/2}$ level.
Apart from these special cases any difference in energy between the excited states due to $m_j'$ dependent Stark or Zeeman shifts will lead to a complex, nonsinusoidal excitation dynamics since $\Omega$ is also dependent on $m_j'$. { Optical pumping of the ground state atoms into the qubit basis, along with a 3 Gauss magnetic bias field, keeps excitation of the undesired Rydberg Zeeman levels to a manageable level.}


\begin{figure}
\centering\includegraphics[width=3.5 in]{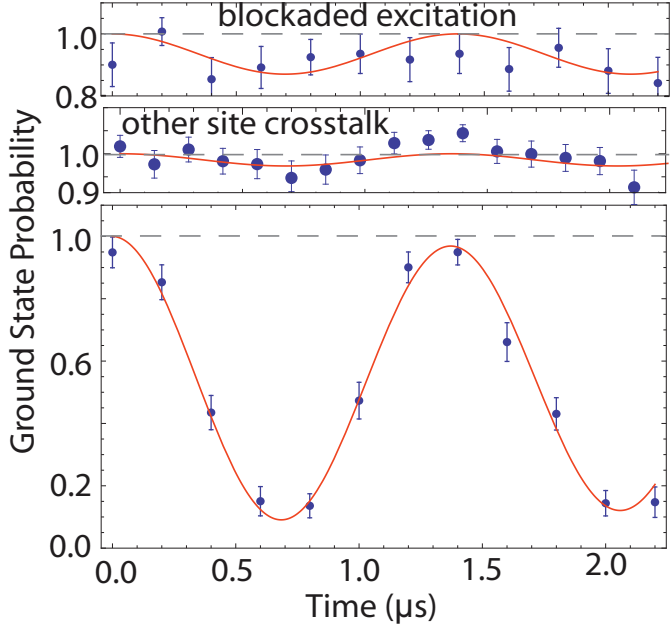}
\caption{ Coherent Rydberg Rabi flopping, and demonstration of blockade.  The bottom frame shows the coherent Rabi flopping of the target qubit observed when the control atom is in the $\ket{0}$ state.  The top frame shows the suppressed flopping of the target atom when the control atom is in the $\ket{1}$ state.  From \protect\cite{Zhang2010}.}\label{fig:blockade}
\end{figure}

\begin{figure}[!t]
\centering
\includegraphics[width=10.cm]{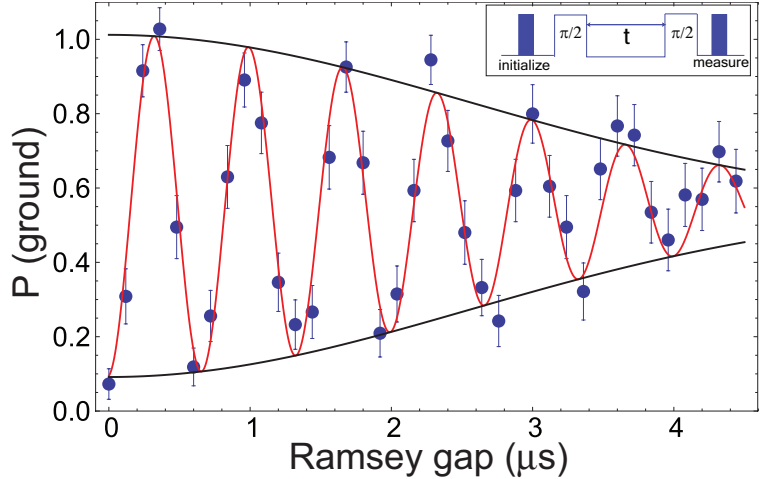}
\caption{Ramsey oscillations between ground and Rydberg states, from \protect\cite{Saffman2011}.
}
\label{fig.rydberg_ramsey}
\end{figure}

With the above considerations taken into account, Rabi oscillations between ground and Rydberg levels have been reported by us and several other research groups   (\cite{Johnson2008,Reetz-Lamour2008b,Miroshnychenko2010,Zhang2010,Zuo2009}).  

Figure \ref{fig:blockade}  shows oscillations of single $^{87}$Rb atoms confined to an optical trap with radius of about $3~\mu\rm m$ and  optically pumped into
$f=2,m_f=0.$ Rydberg excitation to $97d_{5/2},m_j=5/2$ used $\sigma_+$ polarized lasers at 780 and 480 nm. 
The excitation laser beams had waists that were a few times larger than the width of the trapped atom distribution  so the
effects of spatial variation of the Rabi frequency were minimized.
The  FORT laser  was turned off before the excitation lasers were applied. After a variable length excitation pulse the trap was turned on again which photoionized the Rydberg atoms before they could radiatively decay. Loss of a Rydberg atom from the trap therefore provided a signature of successful Rydberg excitation. The less than 100\% observed probability of exciting a Rydberg atom was attributed mainly to Doppler broadening  and the finite Rydberg detection efficiency due to a finite ratio between the photoionization and radiative decay rates.

Further verification of  coherence of the Rabi oscillations was obtained by performing a Ramsey interference experiment. The Ramsey experiment also allows the dephasing rate to be quantified. 
 As shown  in Fig. \ref{fig.rydberg_ramsey} the oscillations decay with a characteristic time of a few $\mu\rm s$. The main factors limiting the coherence were Doppler broadening and magnetic fluctuations (\cite{Wilk2010,Saffman2011}). 
This relatively short coherence time, compared to msec observed for ground state superpositions, limits the fidelity of the Rydberg CNOT gate as will be discussed in the following sections.

\section{Rydberg Blockade} \label{sec:blockadeexpt}

The strong interaction between two Rydberg atoms is the key element for entanglement using Rydberg blockade.  Since at distances of 5-10 microns the atoms are separated by much more than their 1 $\mu$m diameters (at $n=100$), the overlap of the electron wave functions of the two atoms is negligible.  The dominant interaction is then the dipole-dipole interaction
\be
V_{dd}={3( {\bf d}_{1}\cdot {\bf R)(R}\cdot {\bf d}_{2})-{\bf d}_1\cdot {\bf d}_2\over R^3}
\ee
where ${\bf d}_i=-e {\bf r}_i$ is the dipole moment operator for the electron on atom $i$.  Since the dipole operator has odd parity, it mixes in states of opposite parity.  For Rb excited to the $nd$ states, the dominant interaction comes by mixing in states $(n+1)p$ and $(n-1) f$.  The dipole-dipole interaction thus causes transitions
\be
nd+nd\rightarrow (n+1)p+(n-1)f
\ee
Since the $p+f$ state is slightly higher in energy that the $d+d$ state, the direct transfer of population to the $p+f$ state is forbidden, but the dipole-dipole interaction shifts the energies in 2$^{\rm nd}$ order, producing an effective $V_6\sim1/R^6$ van der Waals interaction between the atoms \citep{Walker2008}.  Since we excite only the $m=5/2$ level (referred to a coordinate system aligned with the FORT traps), the van der Waals interaction between the two atoms has a notable amount of variation as a function of the relative positions of the two atoms in the trap.

\begin{figure}
\includegraphics[width=4 in]{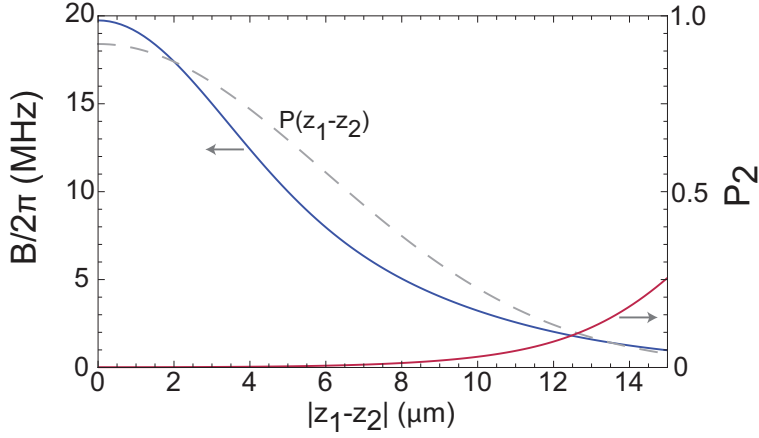}
\caption{Variation of the interaction between two 97d$_{5/2}(m=5/2)$ Rb atoms confined in traps 8.7 $\mu$m apart, as a function of the difference in their relative positions in their individual traps.  The average of $1/B^2$ over the probability distribution $P(z_1-z_2)$ determines the excitation probability $P_2$, shown in red.  From \protect\cite{Zhang2010}}\label{fig:97d}
\end{figure}

To quantify the effectiveness of the dipole-dipole interaction in causing blockade, we note that the probability of double excitation is approximately, for $\Omega_2\ll V_6$,
\be
P_2\approx {\Omega_2^2\over V_{6}^2};\label{p2}
\ee
this is just the probability of off-resonant excitation of the doubly-excited $\ket{rr}$ state \citep{Walker2008}.  For any particular realization of the experiment, the atoms may be as close as the FORT site separation of $8$ microns.  This gives a small excitation probability.  On the other hand, the atoms may also happen to be at the edges of their respective atomic distributions and may approach 15 microns separation, where the dipole-dipole interaction is much weaker.  If the blockade is to fail, it is much more likely to do so when the atoms are as far apart as possible.

The key parameter for blockade effectiveness, then, is not the average strength of the Rydberg-Rydberg interaction, but rather the average of its -2 moment:
\be
{1\over {\sf B}^2}=\int dz_1dz_2 {P(z_1)P(z_2)\over V_6(z_1,z_2)^2}\label{p2b}
\ee
where $P(z)$ is the probability to find the atom at position $z$ inside the FORT trap.
In terms of the blockade shift $\sf B$, Eq.~(\ref{p2}) becomes
\be
P_2\approx {\Omega_2^2\over{\sf B}^2};\label{P2b}
\ee
The interaction potential between two 97d$_{5/2}$ Rb atoms is shown in Fig.~\ref{fig:97d}.  Note that it is plotted as a function of the relative axial trap positions of the two atoms, the relevant coordinates for this type of experiment.

The blockade effectiveness is studied experimentally by exciting one atom via a $\pi$-pulse to the Rydberg state, then performing Rydberg Rabi flopping on the other atom.  This is illustrated in Fig.~\ref{fig:blockade}.  The lower panel shows Rydberg Rabi flopping with no atom in the other trap, while the upper panel shows that with a single atom in the $\ket{1}$ state the Rabi flopping is almost completely suppressed.  {As with ground state rotations, even though the spatial profiles of the excitation lasers partially overlap the adjacent trapping sites, the AC Stark shifts suppress potential cross-talk effects.}

{This successful observation of blockade, by our group and by \cite{Gaetan2009}, was a key milestone in this research, as it was the first direct evidence for a coherent blockade effect between individual atoms.  Many other experiments (reviewed in \cite{Saffman2010}) done in collections of Rydberg atoms had shown evidence for these effects, but not at the two-atom level.  The Gaetan {\it et al.} experiment was also notable for its observation of the predicted $\sqrt{2}$ increase predicted for simultaneous excitation of two blockaded atoms.}

\section{CNOT Gate}\label{sec:CNOT}

\begin{figure}[!t]
\centering
\includegraphics[width=12cm]{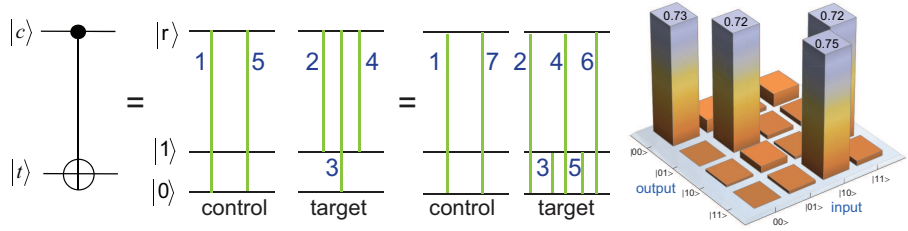}
\caption{(Color online) Amplitude swap CNOT gate pulse sequence  and the experimental result from \protect\cite{Isenhower2010}. All operations are $\pi$ pulses. 
}
\label{fig.as_cnot}
\end{figure}

\begin{figure}[!t]
\centering
\includegraphics[width=12cm]{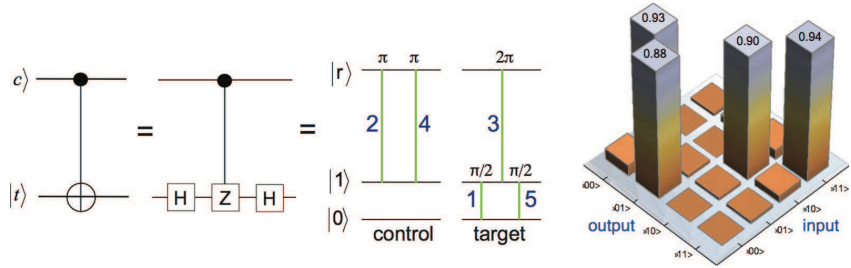}
\caption{(Color online) CNOT gate using Hadamard pulses on the target atom and experimental result from \protect\cite{Zhang2010}. 
}
\label{fig.cnot}
\end{figure}

The Rydberg blockade effect described in the previous section was used to demonstrate a CNOT 
gate (\cite{Isenhower2010,Zhang2010}) which is universal for quantum computation. Two different approaches were successfully realized. In the first, see Fig. \ref{fig.as_cnot}, referred to as an amplitude swap CNOT gate, Rydberg excitation of the control atom blocks a sequence of three pulses which swap the target atom ground states. The control atom is therefore Rydberg excited from $\ket{0}$. 
This form of the gate requires 
only Rydberg pulses, and in principle, no additional ground state rotations. In the actual experimental implementation lasers were not available coupling both $\ket{0}$ and $\ket{1}$ to the Rydberg level. 
Therefore the longer pulse sequence shown in the figure was used. This sequence gives the same  transformation as the ideal CNOT up to single qubit phases. In the second approach, see Fig. \ref{fig.cnot}, $\pi/2$ ground state rotations were applied to the target atom before and after the Rydberg pulse sequence to convert the $C_Z$ gate into a CNOT following the standard circuit construction (\cite{Nielsen2000}).

\begin{figure}[!t]
\centering
\includegraphics[width=12cm]{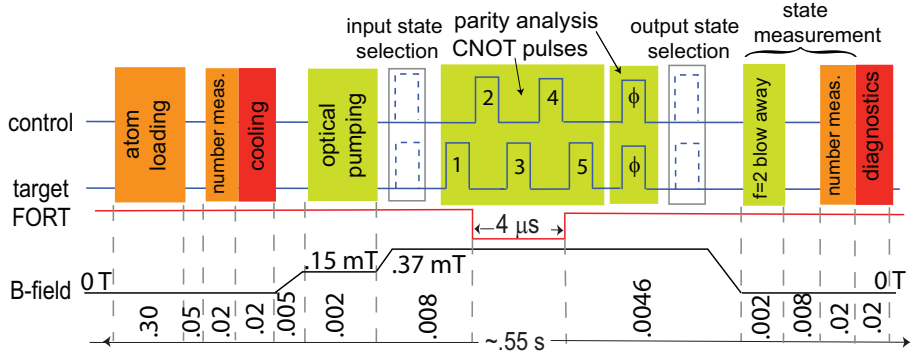}
\caption{Sequence of pulses for CNOT operation.}
\label{fig.exp_sequence}
\end{figure}

The experimental sequence used for the CNOT demonstrations is shown in Fig. \ref{fig.exp_sequence}.  Atoms are first loaded into two sites. Measurements are made to verify the presence of two atoms. If this is not the case the data is discarded. After the first measurement the atoms are recooled with optical molasses and pumped into the $F=2, m_F=0$ upper clock state. Ground state pulses are then applied to select the two-atom input state, the CNOT pulses are applied, an output state is selected using additional ground state pulses, and the output state is measured. 
While the CNOT pulses last only a few $\mu\rm s$ the entire sequence takes about half a second, largely due to the atom loading and measurement steps. 
The box labelled diagnostics at the end of the sequence encompasses  automated measurements used to correct for slow laser intensity drifts.

The measured data shown in Figs. \ref{fig.as_cnot}, \ref{fig.cnot} verify that the gate produces the desired outputs with probabilities ranging from 
0.72 - 0.94. The better performance seen in Fig. \ref{fig.cnot} is not intrinsic, but was due to experimental upgrades in the half year between the two experiments. These upgrades shortened the time between atom loading and CNOT pulses, increased the optical pumping fidelity, and also improved the shot to shot laser stability. The residual errors are primarily due to loss from collisions with untrapped background atoms in the time between the first and second measurements. Additional errors come from imperfect optical pumping, pulse area errors, Doppler broadening, and spontaneous emission from the $5P_{3/2}$ state used in the Rydberg excitation process. The calculated intrinsic error due to imperfect Rydberg blockade and spontaneous emission from the Rydberg states was about 1\%, and was not observable due to the higher level of technical errors. Detailed estimates of the various error sources and comparison with experiment have been given in \cite{Isenhower2010,Zhang2010,Zhang2012}.

Although the probability truth tables look like a CNOT matrix they do not verify that the gate preserves coherence, which is essential for use in quantum logic experiments. Verification of the coherence and entanglement generation capabilities of the gate are discussed in the next section.

\section{Entanglement Verification}\label{sec:parity}

While the population truth tables for the CNOT gate strongly suggest high fidelity quantum manipulation, they are insensitive to dephasing of the coherence between the output states.  The proper way to quantify the effectiveness of the gate is to perform quantum state tomography on the entangled Bell state $\ket{B_1}=\ket{00}+\ket{11}$ produced by action of the CNOT gate on the factorizable input state $\ket{\psi_0}=\ket{0+1}\ket{0}$, or, even better, perform quantum process tomography to evaluate the fidelity of the gate for arbitrary inputs.  Rather than do this, we used Bell-state  population measurements and  measurement of parity oscillations \citep{Sackett2000} to determine the degree of entanglement reached.

\begin{figure}
\includegraphics[width=2.5 in]{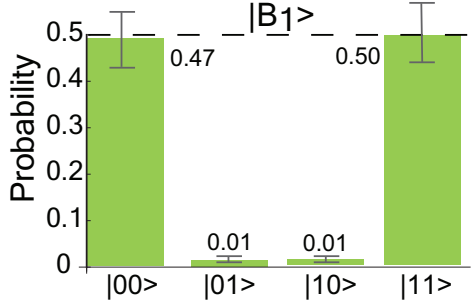}\hspace{0.2 in}
\caption{Measurement of the populations of the Bell state output of the CNOT gate, from \protect\cite{Zhang2010}.}\label{Bellpops}
\end{figure}

Using the $\ket{\psi_0}$ input state, we begin by measuring the probabilities $P_{ij}$ of finding the atoms in the various states $\ket{ij}$; for a Bell state this should give $P_{00}=P_{11}=1/2, P_{01}=P_{10}=0$.  The experimental results are shown in Fig.~\ref{Bellpops},  which indicates a very high degree of correlation between the states of the two atoms.  This is not by itself sufficient to indicate entanglement, as it could represent a non-entangled density matrix
\be
\rho_{\rm ne}\approx{1\over 2}{\rm diag}(1,0,0,0)+{1\over 2}{\rm diag}(0,0,0,1)
\ee
Such a density matrix does not violate Bell's inequalities.  Verification of the Bell state requires measurement of the coherence $\bra{11}\rho\ket{00}$, a task accomplished by measuring parity oscillations \citep{Sackett2000}.

The parity oscillation measurement consists of allowing the output of the CNOT operation to evolve freely for a time $t$ with a frequency shift $\Delta$ with respect to the system clock.  This corresponds to producing the transformation $\ket{0}\rightarrow\ket{0}$ and $\ket{1}\rightarrow e^{-i\phi}\ket{1}$.  Then a $\pi$/2 pulse is applied to each atom, followed by a measurement of the state of each atom.  The parity measurement is $\Pi=P_{00}+P_{11}-P_{01}-P_{10}$. The key insight is that the coherence $\bra{11}\rho\ket{00}$ is mapped to a parity $\Pi=\cos2\phi$ (up to a phase), the coherence $  \bra{10}\rho\ket{01}$ is mapped to a constant $\Pi=1$, and the other coherences have zero parity.  In particular, $\Pi_{B1}=\cos2\phi$.  The parity of $\rho_{\rm ne}$ is zero.

\begin{figure}
\includegraphics[width=5.4 in]{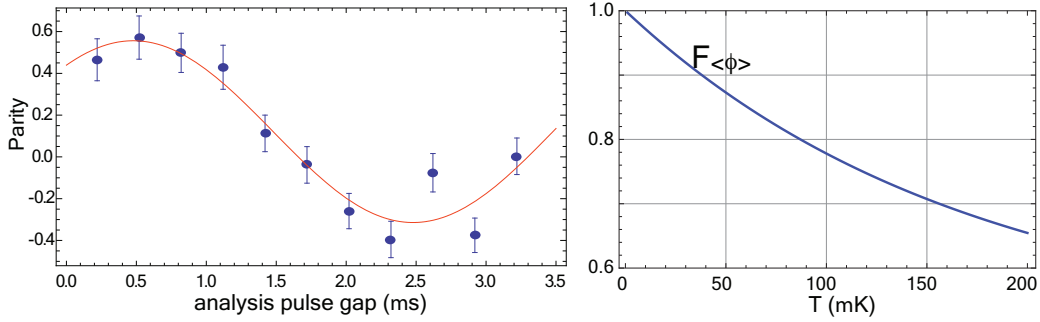}\hspace{0.2 in}
\caption{{Parity oscillations observed to test the Bell-state preparation fidelity, and temperature dependence of the predicted fidelity.  Adapted from \protect\cite{Zhang2010}.}}\label{paritydata}
\end{figure}

The parity measurement is shown in Fig.~\ref{paritydata}.  An important correction that is applied to this data is to account for loss of atoms in the time between the first measurement and the end of the state readout\citep{Zhang2010}.  Since the loss occurs whether or not the gate is operated, we apply an overall correction factor. 

The amplitude of the parity oscillation is $0.44\pm0.03$.  We can use this to estimate the the probability $b$ of obtaining the Bell state:
\be
\rho=(1-b)\rho_{\rm ne}+b\rho_{B1}
\ee
The parity of this density matrix is $b\cos2\phi$, so we conclude that the CNOT gate produces the desired Bell state 44\% of the time.

The usual definition of entanglement fidelity is $F={\rm Tr}(\rho\rho_{B1})$.  Since ${\rm Tr}(\rho_{\rm ne}\rho_{B1})=1/2$, the fidelity with which the Bell state is prepared is
\be
F=(1-b)/2+b=0.72
\ee
It is interesting to note that this commonly used measure of fidelity gives $F=0.5$ when $b=0$; a completely unentangled but classically correlated state has non-zero fidelity by this measure.

We have recently published a detailed analysis of the error sources that contributed to this measurement (\cite{Zhang2012}).  The dominant effect that reduces the parity contrast is dephasing due to the atomic motion, first pointed out by \cite{Wilk2010}.
During the $t=$1.5 $\mu$s time that the control atom is in the Rydberg state, the atomic motion of velocity $v$ results in  a phase change in the two-photon excitation field of $\delta\phi=\delta kvt\sim 1$, thus partially dephasing the coherence of the output state.  These fluctuations are mapped onto the $\ket{11}$ output state
to give
\be
\expect{e^{ikvt}}= e^{-\frac{k^2 t^2 k_BT}{2m}}=0.38
\ee
thus accounting for the observations.  {The predicted fidelity improvement with reducing the temperature is shown in the inset of Fig.~\ref{paritydata}.}

{At nearly the same time as our work, \cite{Wilk2010} demonstrated entanglement using Rydberg blockade by simultaneous excitation of Rydberg states in optically unresolved FORT sites.  Using parity oscillations, they achieved entanglement fidelity of 75\%. }

\section{Future improvements}\label{sec:concl}

We conclude by discussing key new capabilities needed for substantial improvements in the fidelity of Rydberg blockade gates.

\subsection{Deterministic Loading of Optical Lattices}

The experiments that have been discussed here demonstrate the feasibility of Rydberg-mediated quantum information processing at the two-atom level.  Extension of these methods to many atoms is a current challenge.

Trapping in multiple sites for establishing a register of qubits can be accomplished in several ways. Diffractive optical elements, spatial light modulators, and microlens arrays have all been used to split a single beam into arrays of several beams 
(\cite{Bergamini2004,Lengwenus2010,Knoernschild2010}). Optical lattices formed by interference of several beams provide another promising setting
for experiments with many qubits (\cite{Nelson2007,Bakr2009, Sherson2010}).   Although lattices formed from counterpropagating pairs of near infrared beams have sub-micron periods, recent experiments have demonstrated the capability of single site addressing with low crosstalk in such a setting (\cite{Weitenberg2011}). 

Unfortunately even 50\% loading probability, as obtained by loading into tightly confining FORTs \citep{Schlosser2002},  is not adequate for preparation of a large number of qubits since the probability of loading one atom in each of $N$ sites scales as $2^{-N}$. Recent experiments have manipulated the collisional properties of pairs of atoms with blue detuned light to selectively eject single atoms from the trap, instead of a pair (\cite{Grunzweig2010}). This has led to demonstration of 86\% single atom loading probability. Further optimization may be possible (\cite{Grunzweig2011}). 
We proposed  (\cite{Saffman2002}) to use the Rydberg blockade effect to select a single atom from an $N$ atom ensemble and thereby push the single atom loading probability close to 100 \%.  This would have the huge advantage for our work that the protocol can be implemented with essentially no changes in the apparatus, barring optimization for loading tens of atoms into the qubit sites. Preliminary  results have demonstrated 57\% single atom loading and further optimization is in progress.   Recent work \citep{Sortais2012} has shown that for multiple atom loading, light assisted collisions give sub-Poissonian statistics.  Such a sub-Poissonian initial atom number distribution should substantially improve the performance of the deterministic blockade loading scheme.

Optical lattices are also attractive for preparing single atom occupancy of a large number of sites using BEC-Mott insulator transition physics (\cite{Greiner2002}).  While this is an elegant approach for preparing a large quantum register it also adds considerable complexity to an already challenging enterprise if one wishes to use the Mott insulator state as part of a Rydberg state mediated quantum computer. The technical challenges are not fundamental, and will likely be overcome in the near future. 

\subsection{Advantages of dark FORTs}

\begin{figure}
\includegraphics[width=3.5 in]{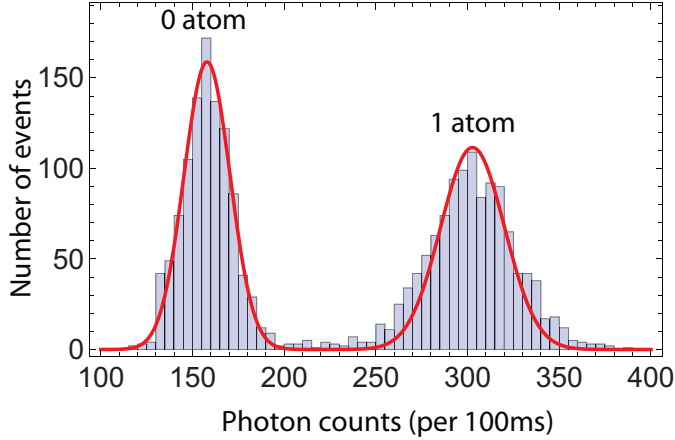}
\caption{Single atom detection in a dark FORT, from \protect\cite{Li2012}. }\label{Lifig}
\end{figure}

Many technical issues facing these experiments may be simultaneously improved by trapping the atoms in blue-detuned FORT light \citep{Li2012}.  The use of such "dark" traps is advantageous for storing and cooling qubits, as the absence of a large intensity at the location of the trapped atom reduces light shifts which otherwise interfere with the most straightforward laser cooling mechanisms, and also removes the need to detune qubit readout light to account for trap light induced shift of the cycling frequency. There is also less light scattering in a dark optical trap which improves coherence times although,  surprisingly, a detailed calculation shows that the  mechanical heating rate of sufficiently cold atoms is the same for red and blue detuned traps (\cite{Gerbier2010}). Bottle beam  configurations have been used to trap small ensembles (\cite{Isenhower2010}), and also single atoms
(\cite{Xu2010,Li2012}) with good localization  suitable for qubit experiments.  Similar improvements in readout are expected, and were recently reported by \cite{Li2012}, with a single-atom detection histogram shown in Fig.~\ref{Lifig}.

In addition, since Rydberg states have negative polarizabilities, it may be possible to realize traps where the Rydberg excitation frequency is unaffected by the presence of the light, allowing the Rydberg excitations to be accomplished without temporarily shutting off the FORT light.

In order to have the same trapping potential for ground and Rydberg atoms the polarizability must be the same for both states. We must therefore choose a wavelength for which the ground state polarizability is negative, and use either a 3D lattice or dark optical trap which captures the atom at a local  minimum of the intensity (\cite{Friedman2002}). In Rb and Cs the ground state polarizability is negative in a broad region to the blue of the first resonance lines. Although there are specific wavelengths for which the ground and Rydberg polarizabilities match exactly, this does not give equal trap potentials. The wavefunction of the Rydberg electron for states with $n\sim 100$ is spread out over a $\mu\rm m$ sized region and it is therefore necessary to average the light shift over the extent of the wavefunction. As was first pointed out by \cite{Dutta2000} this gives an effective ponderomotive potential 
\be
U_R=-\frac{\alpha_e}{2\epsilon_0 c} \int d^3r\, I({\bf R}+\bfr)|\psi({\bf R};\bfr)|^2.
\ee
Here $\bf R$ is the center of mass atomic position and $\psi$ is the electron wavefunction. Since the Rydberg wavefunction is strongly delocalized compared to the ground state matching of the trapping potential in a dark trap occurs when $|\alpha_e|<|\alpha_g|$. Calculations show (\cite{SZhang2011}) that it is possible to achieve perfect matching at trap center for specific Rydberg states and trap geometries. The matching is only approximate when we include the effect of atomic motion at finite temperature, but the mismatch can be limited to a  small fraction of the trap depth provided the atoms are sufficiently cold. Recent experiments from the Raithel group have demonstrated the presence of ponderomotive potentials for Rydberg atoms (\cite{Younge2010}) as well as Rydberg state trapping (\cite{Anderson2011}). Also electrodynamic and magnetic traps are possible candidates for Rydberg trapping (reviewed in \cite{Saffman2010}) although integrating them with other aspects of the total experimental apparatus shown in Fig. \ref{fig.setup} would present new challenges. 

An additional concern with the use of optical traps for Rydberg states is photoionization. This is strongly suppressed for $S$ state atoms near the center of a dark trap due to parity, and because the matrix elements for photoionization of $S$ states are several orders of magnitude smaller than for $P$ or $D$ states (\cite{Saffman2005a}).
Even for the higher angular momentum states the photionization rates in dark traps are substantially smaller than the rates for radiative decay to lower levels.

\subsection{Two-photon excitation via the alkali 2nd resonance}

A way of mitigating the power requirements of two-photon excitation via the first resonance is to use excitation via the second resonance line, $6P$ in Rb or $7P$ in Cs. For Rb this requires wavelengths of  421 and 1014 nm. More power is needed at 421 nm than would otherwise be required at 780 nm, but a very modest $100~\mu\rm W$ is still more than adequate. It is easier to generate high power at 1014 than 480 nm, and in addition the combination of a larger matrix element to the Rydberg level and a smaller spontaneous decay rate from the $p$ level, make for a substantial reduction in power needs. This approach has been demonstrated in \cite{Viteau2010,Viteau2011}. It should be mentioned that there is a principal difference in working with the second resonance line in that two-photon absorption of the first laser will ionize the atom. It is therefore necessary that the rate for this process be small compared to the coherent excitation rate. Since  the energy of two $\omega_1$ photons is far above the ionization threshold the matrix element is small compared to that of the targeted process (\cite{Gallagher1994}), and this should not be a serious problem.

\subsection{Improved FORT decoherence}\label{sec:FORTdecoh}

In addition to photon scattering, qubits in optical traps suffer decoherence due to higher multipoles of the polarizability tensor which give $F$ and $m_F$ dependent energy shifts. The differential light shifts scale as (\cite{Kuhr2005}) $\delta U\sim U_F \frac{\omega_{\rm hf}}{\Delta}$. Magnetic field offsets or fluctuations also give rise to differential energy shifts. A large body of recent work has shown that by careful selection of the FORT detuning, optical polarization, magnetic field strength and direction, and atomic Zeeman state it is possible to cancel both magnetic and light shifts (\cite{Flambaum2008,Derevianko2010a,Derevianko2010b,Lundblad2009,Chicireanu2011,Dudin2010,Radnaev2010}). A particularly promising and flexible approach has been demonstrated in \cite{Radnaev2010} which uses two-photon dressing to cancel trap induced light shifts. Coherence times exceeding 100 ms were achieved in an  atomic ensemble, and even better performance can be expected with single atoms.

\subsection{Fundamental limits}

Assuming continued technical progress, 
 we have recently analyzed in  detail  the physical limits on the performance of Rydberg CNOT-based quantum computing  \citep{Zhang2012}.  Fundamentally, the process is limited by the spontaneous emission lifetime of the Rydberg state and the strength of the blockade interaction.  These parameters seem to lead to the lowest fundamental error rates when $n\sim100$ Rydberg states are used.  This study suggests that fundamental error probabilities below $0.002$ are possible.

It is interesting to note that there are many other applications of Rydberg blockade being pursued. Many of these were recently reviewed \citep{Saffman2010}. These include non-linear optics, entanglement of ensembles, dressing of BEC's with Rydberg states in order to produce controllable anisotropic interactions, and generation of novel quantum states of light.  The success of the two-atom experiments discussed in this paper bodes well for many of these other possibilities as well.

\section{Acknowledgements}

This work was accomplished by a number of talented students and post-docs.  These include Todd Johnson, Erich Urban, Nick Proite, Pasad Kulatunga, Thomas Henage and Deniz Yavuz in the early phases of the work, building up from  scratch to the first CNOT studies.  More recently, Larry Isenhower, Alex Gill, and Xianli Zhang performed the entanglement demonstrations.  This work has been supported  by the National Science Foundation, with  additional funding from the ARO, DTO, ARDA, IARPA, and DARPA.




\bibliographystyle{elsart-harv}

\bibliography{RydAdv}

\end{document}